\documentclass[epjc,preprint]{revtex4}%
\usepackage{amsfonts}
\usepackage{amsmath}
\usepackage{amssymb}
\usepackage{graphicx}%
\begin{document}
\title{Generation of entanglement between two laser pulses through gravitational interaction}
\author{Feifan He$^{a,b}$}
\author{Baocheng Zhang$^{b}$}
\email{zhangbaocheng@cug.edu.cn}
\affiliation{$^{a}$Institute of Geophysics and Geomatics, China University of Geosciences,
Wuhan 430074, China}
\affiliation{$^{b}$School of Mathematics and Physics, China University of Geosciences,
Wuhan 430074, China}
\begin{abstract}
We investigate entanglement generation between two pulses through the
gravitational interaction in the framework of linearized quantum gravity.
Different from the earlier suggestions that two massive particles can be
entangled through the gravitational field produced by the particles
themselves, we use the massless particles (a laser pulse) to generate the
gravitational field. In our proposal, the propagating retarded effect of the
gravitational perturbation generated by the laser pulses can be incorporated
into the interaction, which is significant for supporting the assumption of
local interaction. It is found that the locally linearized quantum
gravitational interaction can indeed lead to the generation of entanglement
between two pulses, and the result is dependent on the propagating time of the
gravitational perturbation. We also provide a measurement suggestion using a
Holometer-like interferometers structure to present this result of
gravitation-induced entanglement, which could show that the gravitational
mediator is quantum in future experiments.

\end{abstract}
\maketitle

\section{Introduction}

The most fundamental physical theories, quantum theory and general relativity,
claim to be universally applicable and have been confirmed to have high
accuracy in their respective domains. However, it is hard to merge them into a
unique corpus of laws. Quantizing gravity is one of the most intensively
pursued areas of physics \cite{od09}. Some approaches are based on applying a
quantization procedure to the gravitational field, in analogy with that made
for the electromagnetic field \cite{tbc82}; some are based on
\textquotedblleft geometrizing\textquotedblright\ quantum physics \cite{sg98};
others modify both quantum physics and general relativity into a more general
theory (e.g. string theory \cite{lt89}) but contains them as special cases
\cite{kc06,rc08}. However, the lack of empirical evidence for the quantum
aspects of gravity has led to a debate on whether gravity is a quantum entity.

Recently, two groups of scientists proposed detectable methods in laboratory
\cite{mv170,bmm17,mv17}, in which they showed that the generation of
entanglement between two mesoscopic test masses in adjacent matter-wave
interferometers could be realized to confirm the quantum character of the
gravitational interaction. This opened up an alternative direction for
exploring the quantum aspect of gravity again
\cite{mvd18,mv18,bcm18,cbu19,cr19,ktp20}. Even so, there are still some
suspicious discussions \cite{hr17,ah18} on this criterion due to the
gravitational interaction between the particles being presented in a direct
and instant way. In the answers \cite{mv19} to these questions, a crucial
element about the assumption of locality is emphasized. In particular, the
criterion to ensure the locality \cite{bwa18} had also been analyzed. A
detailed discussion about the necessity for the locality of the gravitational
interaction in the framework of linearized quantum gravity was made to confirm
the quantum property of linearized gravity \cite{mmb20}. In this paper, we use
a practical example to present the influence of locality assumption by
incorporating the finite-speed propagation of the gravitational field into the
result of entanglement generation.

Since the gravitational wave had been tested with the finite speed
\cite{abp16}, we will consider the gravitational interaction with the form
similar to the gravitational wave and present its finite-speed propagation by
the retarded effect. As well-known, it is hardly possible to generate the
observable gravitational wave in the lab \cite{rb06}. However, an interesting
proposal for the propagating gravitational field generated by light
\cite{tep31,bwb69,as71,smo79,kd98,jb06,bw09,al14,rwm16} has caused much
attention in the past years. In this paper, we will investigate whether the
time-varying gravitational field generated by the laser pulses can lead to the
generation of entanglement between two laser pulses. We will also calculate
the quantity of entanglement generated between two laser pulses by the
gravitational field. Based on these, our work presents a significative
extension for the earlier suggestions about the quantum property of the
gravitational field by using massless particles, while nearly all the earlier
methods used massive particles. At the same time, our work can support the
assumption of locality required to confirm the quantum property of the
gravitational field to a certain extent, but it cannot solve all the
disputations in this field, which has to be ended until decisive experimental
results appear.

In order to realize our thought experimentally, we consider two coupled
interferometers, similar to the earlier suggested \textquotedblleft
Holometer\textquotedblright\ \cite{hcj12} which has been implemented at Fermi
lab \cite{cjt16} for better studying the quantum character of spacetime. By
properly setting up two interferometers, we will investigate the mutual
influence of the pulses propagating in the two interferometers through the
gravitational field generated by them, and compare the results of the
interference in a single interferometer for the case that the other
interferometer is placed nearby and for the case that there is no other interferometer.

The paper is organized as follows. In the second section, we will review and
recalculate the gravitational field produced by a laser pulse using the
earlier method given by M. O. Scully \cite{smo79}. Then, we discuss the
gravitational interaction between two pulses and compare it with Newton's
gravitational law in the third section. At the same time, the retarded effect
of propagating the gravitational perturbation generated by the pulses is also
considered in the interaction. In the fourth section, we discuss the change of
the quantum states in the interferometers and calculate the change of
entanglement due to the gravitational interaction. We also investigate whether
and how entanglement can be produced. Finally, we summarize the results in the
fifth section.

\section{Gravitational field produced by a laser pulse}

As well-known, the metric of spacetime can be influenced by the energy density
according to general relativity. If the energy is not large enough, the change
of the spacetime metric is small. Thus, when a high-power laser pulse moves in
a flat spacetime, the metric becomes%
\begin{equation}
g_{\mu\nu}=\eta_{\mu\nu}+h_{\mu\nu},
\end{equation}
where the flat spacetime metric is given by $\eta_{\mu\nu}=\left(
c^{2},-1,-1,-1\right)  $ and $h_{\mu\nu}$ represents the perturbation of the
metric caused by the laser pulse. According to the Einstein field equation,
the small correction $h_{\mu\nu}$ obeys the linearized equation \cite{mtw73}%
\begin{equation}
\square^{2}h_{\mu\nu}=\kappa(T_{\mu\nu}-\frac{1}{2}\eta_{\mu\nu}T),
\label{leq}%
\end{equation}
in which the constant $\kappa$ is related to the gravitational constant $G$,
given by $\kappa=16\pi G/c^{2}$. $T_{\mu\nu}=\varepsilon_{0}\left(  F_{\mu
}^{\lambda}F_{\lambda\nu}-\frac{1}{4}\eta_{\mu\nu}F^{\sigma\rho}F_{\sigma\rho
}\right)  $ is the stress-energy tensor \cite{jjd07} of electromagnetic field
from the pulse, and $T=\eta^{\mu\nu}T_{\mu\nu}$ is the trace of the tensor
$T_{\mu\nu}$.

Consider that the laser propagates along the $x$ direction (see Fig. 1 for the
coordinates) with a velocity $v$ ($v<c$) since the gravitational perturbation
cannot propagate along $y$ or $z$ direction (see Eq. (\ref{metric-tensor2})
below) when $v=c$. This can be realized \cite{smo79} by guiding the pulse to
travel through a material medium with the index of refraction $n>1$. Thus, the
field carried by the laser pulse is given as \cite{kb72},%
\begin{equation}
E_{2}\left(  x,t\right)  =\epsilon\left(  \overrightarrow{r},t\right)
\sin\left(  \omega t-kx\right)  \text{,}%
\end{equation}%
\begin{equation}
B_{3}\left(  x,t\right)  =\frac{v}{c}\frac{\epsilon\left(  \overrightarrow
{r},t\right)  }{c}\sin\left(  \omega t-kx\right)  \text{,}%
\end{equation}%
\begin{equation}
B_{1}\left(  x,t\right)  =[1-(\frac{v}{c})^{2}]^{\frac{1}{2}}\frac
{\epsilon\left(  \overrightarrow{r},t\right)  }{c}\cos\left(  \omega
t-kx\right)  \text{,}%
\end{equation}
where $\epsilon\left(  \overrightarrow{r},t\right)  $ denotes the envelope of
the laser pulse moving with velocity $v$, and its expression is given as
$\epsilon^{2}\left(  \overrightarrow{r},t\right)  =E_{0}^{2}A[\theta\left(
v\left(  t+t_{d}\right)  -x\right)  -\theta\left(  vt-x\right)  ]\delta\left(
y\right)  \delta\left(  z\right)  $ for the duration $t_{d}$ of the pulse, in
which the spatial variations in the $y$ and $z$ directions have been ignored,
$E_{0}$ is the laser amplitude, $A$ is the effective cross-sectional area, and
$\theta\left(  x\right)  $ is the step function defined by $\theta\left(
x\right)  =1$ for $x\geq0$; $\theta\left(  x\right)  =0$ for $x<0$. Therefore,
the stress-energy tensor becomes%
\begin{equation}
T_{\mu\nu}=\frac{1}{2}\varepsilon_{0}\epsilon^{2}\left(  \overrightarrow
{r},t\right)  M_{\mu\nu} \label{see}%
\end{equation}
where%
\begin{equation}
M_{\mu\nu}=\left[
\begin{array}
[c]{cccc}%
1 & -\frac{v}{c^{2}} & 0 & 0\\
-\frac{v}{c^{2}} & \frac{v^{2}}{c^{4}} & 0 & 0\\
0 & 0 & 0 & 0\\
0 & 0 & 0 & \frac{1}{c^{2}}[1-\left(  \frac{v}{c}\right)  ^{2}]
\end{array}
\right]  \text{.}%
\end{equation}
Note that the lifetime of one laser pulse in the interferometer is very short,
so we can average the energy density by continuously emitting a bunch of
pulses which have high energy or high frequency \cite{smo79}. Thus, we can
replace the high frequency terms such as $\sin^{2}\omega t$ with $\frac{1}{2}%
$, and neglect the terms as $\frac{1}{2}\sin\omega t$ since they would average
to zero in a short time. It is easy to confirm that the trace $T$ of stress
energy tensor is zero from Eq. (\ref{see}). Hence, the linearized equation
(\ref{leq}) becomes%
\begin{equation}
\square^{2}h_{\mu\nu}=\kappa T_{\mu\nu}. \label{leq2}%
\end{equation}
By solving this equation, one obtains%
\begin{equation}
h_{\mu\nu}=h\left(  \overrightarrow{r},t\right)  M_{\mu\nu}\text{,}
\label{metric-tensor1}%
\end{equation}
where%
\begin{equation}
h\left(  \overrightarrow{r},t\right)  =-\frac{4G\rho A}{c^{2}}\ln\left(
\frac{v\left(  t+t_{d}\right)  -x+\left[  \left(  v\left(  t+t_{d}\right)
-x\right)  ^{2}+(1-v^{2}/c^{2})\left(  y^{2}+z^{2}\right)  \right]  ^{1/2}%
}{vt-x+\left[  \left(  vt-x\right)  ^{2}+(1-v^{2}/c^{2})\left(  y^{2}%
+z^{2}\right)  \right]  ^{1/2}}\right)  . \label{metric2}%
\end{equation}
For a short pulse (i.e. the length of the laser pulse is much smaller than the
arm length of the interferometer) that satisfied the relation, $vt_{d}%
/[(x-vt)^{2}+(1-v^{2}/c^{2})\left(  y^{2}+z^{2}\right)  ]^{1/2}\ll1$, the Eq.
(\ref{metric2}) becomes%
\begin{equation}
h\left(  \overrightarrow{r},t\right)  =\frac{-4G\rho V}{[(x-vt)^{2}%
+(1-v^{2}/c^{2})\left(  y^{2}+z^{2}\right)  ]^{1/2}c^{2}}
\label{metric-tensor2}%
\end{equation}
where the radiation energy density $\rho$ is given as $\rho=\frac{1}%
{2}\varepsilon_{0}E_{0}^{2}$, and the volume of the pulse is $V=Avt_{d}$.
Thus, we obtain the gravitational field generated by a moving laser pulse.

\section{Gravitational interaction between two pulses}

In this section, we will discuss the gravitational interaction between two
pulses in the frame of linearized quantum gravity and show that this
interaction is quantum. To distinguish them from those presented in the last
section, we add hats for the quantum operators, i.e. $h_{\mu\nu}$ is written
as ${\hat{h}}_{\mu\nu}$.

As presented in Ref. \cite{smo79}, the perturbation of the spacetime metric
could influence the motion of the nearby probe pulse and lead to a change in
its amplitude and phase. In that paper, the probe pulse is much weaker than
the laser pulse that generates the gravitational field. In this paper, we will
discuss the interaction between two laser pulses that has equivalent high
power. For this purpose, we consider such gravity-matter coupling interaction
Hamiltonian \cite{rb06},%
\begin{equation}
{\hat{H}}_{int}^{G}=-\frac{1}{2}\int{\hat{h}}_{\mu\nu}^{A}{\hat{T}}_{B}%
^{\mu\nu}d^{3}r-\frac{1}{2}\int{\hat{h}}_{\mu\nu}^{B}{\hat{T}}_{A}^{\mu\nu
}d^{3}r,\label{hamil-int}%
\end{equation}
where $A$ and $B$ are the signs of two pulses, respectively, ${\hat{T}}%
^{\mu\nu}$ is the stress-energy tensor of the pulse, ${\hat{h}}_{\mu\nu}$ is
the quantum perturbation of the metric tensor generated by the laser pulse,
and the range of integral is the length of the pulse. For the first term in
Eq. (\ref{hamil-int}), it indicates that pulse $B$ is influenced by ${\hat{h}%
}_{\mu\nu}$ generated by pulse $A$, and the second term represents pulse $A$
being influenced by ${\hat{h}}_{\mu\nu}$ generated by pulse $B$. For
simplicity, we assume here that the two pulses are the same (i.e. the initial
phase and amplitude are the same), are emitted at the same time, and propagate
through the same dielectric medium. Thus, the two terms in Eq.
(\ref{hamil-int}) have the same expression. Then we build such a coordinate
system that the two pulses propagate along the $x$-axis and are separated from
each other with the distance $d$ ($d\gg vt_{d}$) along the $y$-axis for the
whole propagating process, as presented in Fig. 1, but the pulses are emitted
simultaneously here. This would make our following discussions constrained to
$x-y$ plane. When the velocity of two laser pulses is the same, and the
relative distance between them is $d$, by taking $d=\sqrt{1-v^{2}/c^{2}}y$ and
$D=\sqrt{(x-vt)^{2}+d^{2}}$, the interaction Hamiltonian becomes
\begin{equation}
{\hat{H}}_{int}^{G}\simeq\frac{4G\rho^{2}V^{2}}{Dc^{4}}%
,\label{Hamiltonian-laser}%
\end{equation}
which is obtained using the quantum method from Eq. (\ref{hamil-int}) as
presented in Appendix2. It is noticed that $\rho V$ is the energy of the laser
pulse. To understand this Hamiltonian better, we write the effective mass of
the pulse as $m_{eff}=\rho V/c^{2}=E_{laser}/c^{2}$, and the similar Newtonian
result is%
\begin{equation}
H_{int}^{G}=\frac{4Gm_{laser}m_{eff}}{D}=4m_{eff}\phi_{G}%
.\label{Hamiltonian-laser-field}%
\end{equation}
where the prefactor $4$ $(=2\times2)$ is derived partly (a factor $2$) from
the fact that the general-relativistic result is roughly twice that obtained
from Newtonian theory and partly (another factor $2$) from the same two terms
in Eq. (\ref{hamil-int}). In particular, an evident difference is that
$\phi_{G}$ is time-dependent and similar to a retarded Newtonian potential. If
the initial position is taken as $x=0$ when time $t=0$ and two pulses are
emitted simultaneously, the influence of the gravitational perturbation
generated from one pulse on another pulse occurs at $x=vt$. Thus, $D=d$, and
it seemed that the potential $\phi_{G}$ is time-independent. However, the
propagation of gravitational perturbation is hidden in the moving $x$
coordinates. The intrinsic reason is that two pulses with the initial
positions taken as $x=0$ are emitted simultaneously. It has a similar effect
to the \textquotedblleft gravitational wave\textquotedblright, i.e. it can
propagate, as presented through the retarded influence of the gravitational
perturbation on another pulse (the detailed calculation is given in the appendix).

\begin{figure}[ptb]
\centering
\includegraphics[width=1\columnwidth]{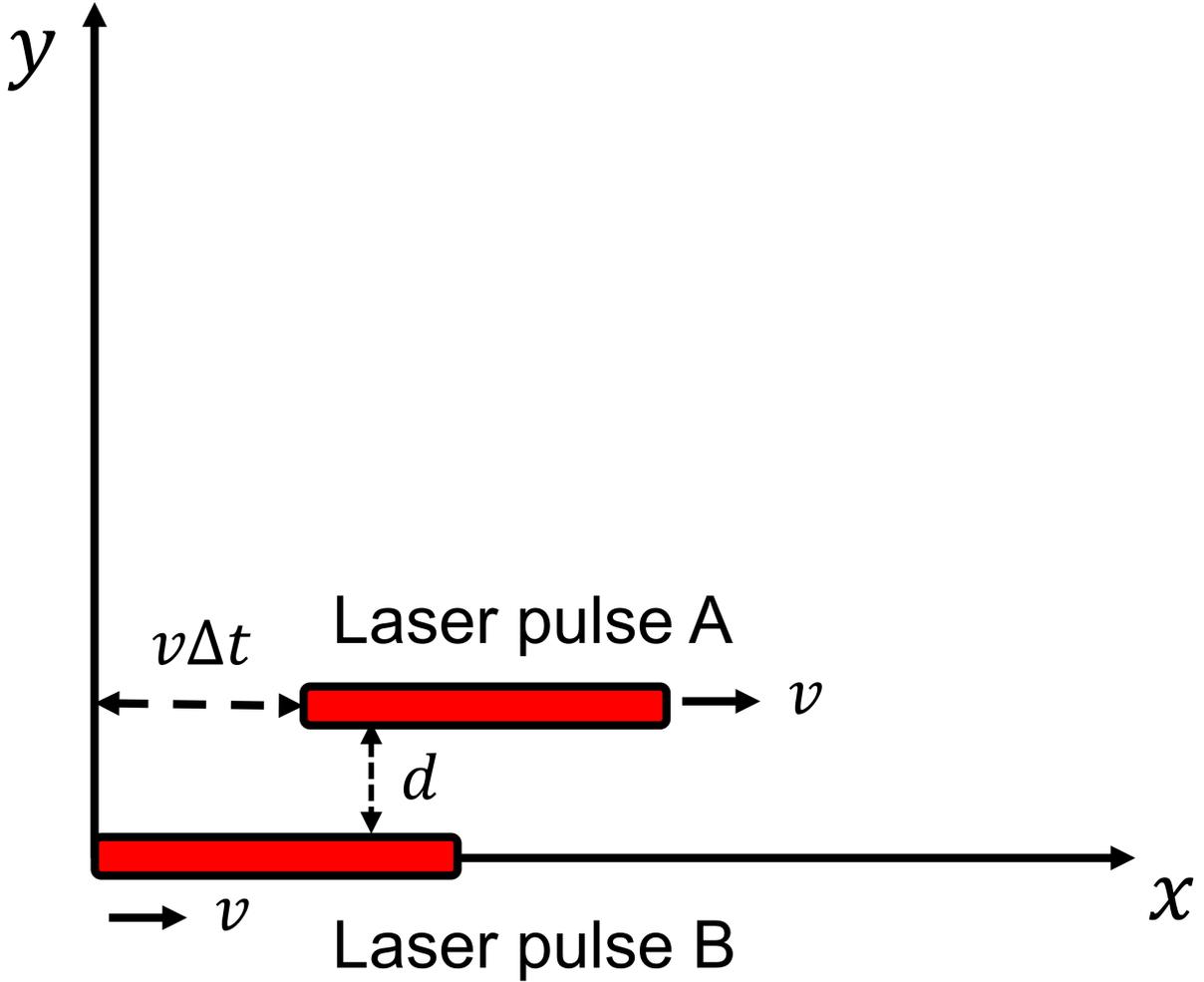} \caption{(Color online) The
diagram for the two pulses with the coordinates. Two pulses propagate along
the $x$ direction but with different initial $y$ values. Pulse $A$ is emitted
earlier than pulse $B$ with a time different $\Delta t$. }%
\label{Fig1}%
\end{figure}

In order to show the retarded effect, we can emit the two pulses with
different initial times, i.e. pulse $B$ is emitted later than pulse $A$, with
a time difference $\Delta t$, as presented in Fig. 1. It requires that
$t_{d}<\Delta t<d/c$ where we assume that the propagating velocity of the
gravitational perturbation is the same as the speed of light in vacuum. The
first inequality ensures that the retarded effect can be shown properly
without overlap along the $y$-direction. The latter inequality ensures that
the influence of the gravitational perturbation can propagate to the other
pulse. The two pulses are not emitted simultaneously, so the two terms in Eq.
(\ref{hamil-int}) are slightly different. Because pulse $A$ is emitted
earlier, its gravitational influence can reach the position where pulse $B$
would be emitted. When pulse $B$ is emitted, it instantly feels the
gravitational influence from pulse $A$. Thus, the first term in Eq.
(\ref{hamil-int}) is obtained as ${\hat{H}}_{int,AB}^{G}\simeq\frac{2G\rho
^{2}V^{2}}{\sqrt{(x_{B}-vt)^{2}+d^{2}}c^{4}}\overset{x_{B}\left(  0\right)
\rightarrow0}{\longrightarrow}$ $\frac{2G\rho^{2}V^{2}}{dc^{4}}$, where
$x_{B}\left(  0\right)  \rightarrow0$ means taking the limit of $x_{B}$ going
to zero for $t$ going to zero. But when pulse $B$ is emitted, pulse $A$ cannot
feel the gravitational influence generated by pulse $B$ instantly and has a
retardance with time difference $\Delta t$. So, we have ${\hat{H}}%
_{int,BA}^{G}\simeq\frac{2G\rho^{2}V^{2}}{\sqrt{(x_{A}-vt)^{2}+d^{2}}c^{4}%
}=\frac{2G\rho^{2}V^{2}}{\sqrt{(x_{B}-vt+v\Delta t)^{2}+d^{2}}c^{4}}%
\overset{x_{B}\left(  0\right)  \rightarrow0}{\longrightarrow}\frac{2G\rho
^{2}V^{2}}{\sqrt{\left(  v\Delta t\right)  ^{2}+d^{2}}c^{4}}$. This leads to
the total interaction Hamiltonian as,%
\begin{equation}
{\hat{H}}_{int,r}^{G}={\hat{H}}_{int,AB}^{G}+{\hat{H}}_{int,BA}^{G}%
\overset{x_{B}\left(  0\right)  \rightarrow0}{\longrightarrow}\frac{2G\rho
^{2}V^{2}}{dc^{4}}+\frac{2G\rho^{2}V^{2}}{\sqrt{d^{2}+\left(  v\Delta
t\right)  ^{2}}c^{4}}.\label{HL2}%
\end{equation}
The Hamiltonian can present the retarded effect to a certain extent. It has to
be pointed out that the time in the expression (\ref{HL2}) is not the time for
the interaction between the pulse and the gravitational field; it referred to
the propagating time of the metric perturbation from the active pulse
(producing the perturbation) to the passive pulse. When $\Delta t=0$, the
expression of the interaction (\ref{HL2}) decays into Eq.
(\ref{Hamiltonian-laser}). It is just the case in which the two pulses are
emitted simultaneously. For presenting the retarded influence evidently, one
can calculate the metric perturbation in Eq. (\ref{leq2}) with the retarded
solution, and see Appendix1 for the details. We find that the same result as
Eq. (\ref{HL2}) is obtained using the selected parameters. The retarded effect
is significant for our result. When we assume that pulse $B$ is emitted later
than pulse $A$, the interaction happens later than the case in which the two
pulses are emitted simultaneously. Since the gravitational interaction has a
finite speed as well-known but this can not be embodied only using the
Newtonian form of gravity, our result represents a self-consistency check for
the locality of the interaction. It is crucial for the confirmation of the
quantum property of the gravitational field and will be discussed in the
following section.

\section{Generation of entanglement}

As presented in Ref. \cite{smo79}, the gravitational field generated by the
active laser pulse can influence the traveling time of the probe pulse
propagating parallel to the active one. This would also result in a different
phase shift for the probe pulse. But the phase shift is induced by the
interaction with a classical gravitational field without requiring the quantum
gravitational field. So, to show that the gravitational field is quantum, one
would have to show that the gravitational field is capable of existing in a
quantum superposition state. In other words, if the local gravitational
interaction can generate entanglement between two pulses that are initially
independent of each other, the mediator of the interaction must possess some
quantum features \cite{bmm17,mv17,mmb20}. Although the two pulses, as stated
in the last section, have the gravitational interaction (\ref{HL2}),
entanglement cannot be generated directly. A beam splitter (BS) has to be
added, as presented in Fig. 2.

\begin{figure}[ptb]
\centering
\includegraphics[width=1\columnwidth]{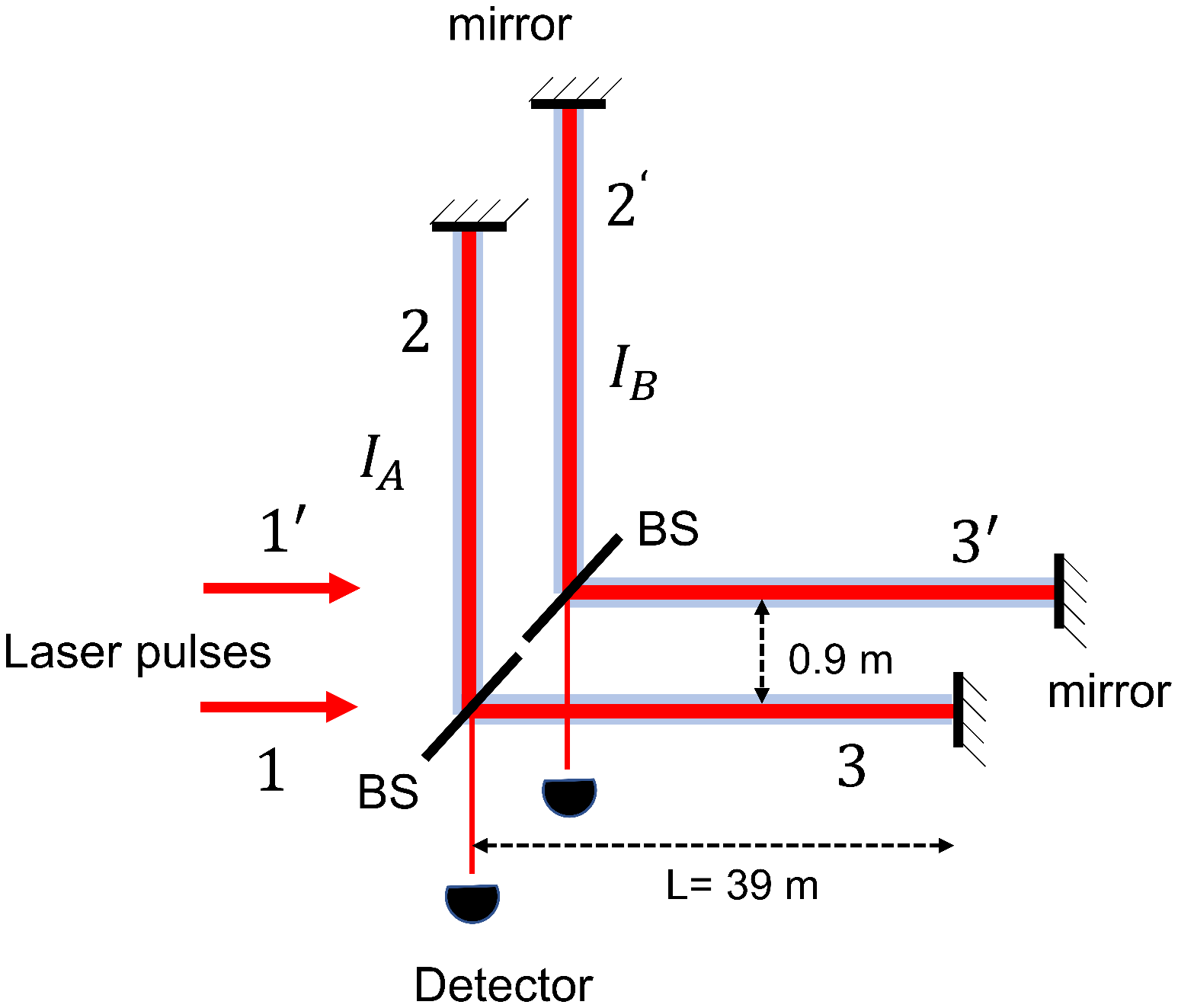} \caption{(Color online) The
diagram for the generation of entanglement between two pulses using the
Holometer-like structure of two interferometers. Two pulses are guided through
the respective beam splitters (BS). Two BSs are placed parallelly, which makes
the paths parallel between $2$ and $2^{\prime}$, $3$ and $3^{\prime}$,
respectively. The mirrors are used to make the pulses reflected and pass
through the BSs for the second time, which makes the final pulses interfere at
the exit. All paths that the pulses move through are horizontal at the same
height and full of the medium with the index of refraction $n>1$. Due to the
existence of the second interferometer $I_{B}$, the result of the interference
of the first interferometer $I_{A}$ would be influenced. }%
\label{Fig2}%
\end{figure}

Now we see how entanglement is generated by the addition of the BS. Assume
that two pulses are in the coherent states respectively, and each coherent
state $\left\vert \alpha\right\rangle ^{k}$, $k=A,B$, passing through the Kerr
medium and a BS \cite{gv08}, becomes the superposition states $\frac{1}%
{\sqrt{2}}\left(  \left\vert -\alpha\right\rangle _{2}^{k}+i\left\vert
\alpha\right\rangle _{3}^{k}\right)  $ where the minus sign represents the
addition of a $\pi$ phase and the imaginary number $i$ represents the addition
of a half-$\pi$ phase. After passing the BSs, the whole state for the two
pulses can be written as%
\begin{equation}
\left\vert \Psi\left(  t=0\right)  \right\rangle ^{AB}=\frac{1}{\sqrt{2}%
}\left(  \left\vert -\alpha\right\rangle _{2}^{A}+i\left\vert \alpha
\right\rangle _{3}^{A}\right)  \frac{1}{\sqrt{2}}\left(  \left\vert
-\alpha\right\rangle _{2^{^{\prime}}}^{B}+i\left\vert \alpha\right\rangle
_{3^{^{\prime}}}^{B}\right)  \text{,}%
\end{equation}
where the gravitational interaction between two pulses before entering the
beam splitters is ignored since it only contributes a common phase factor that
is unobservable. When the laser pulses move in the two perpendicular paths,
the gravitational interaction between pulses along paths $2$ and $3^{\prime}$,
or along paths $3$ and $2^{\prime}$ is ignored since it is weaker than the
interaction between the pulses along paths $2$ and $2^{\prime}$, or along
paths $3$ and $3^{\prime}$. Thus, the final state for the two pulses at the
detectors can be given as
\begin{align}
&  \left\vert \Psi\left(  t=\frac{2L}{v}\right)  \right\rangle ^{AB}%
\nonumber\\
&  =\frac{1}{2}\left(  e^{-i\phi_{22^{^{\prime}}}}\left\vert \alpha
\right\rangle _{2}^{A}\left\vert \alpha\right\rangle _{2^{^{\prime}}}%
^{B}-i\left\vert \alpha\right\rangle _{2}^{A}\left\vert \alpha\right\rangle
_{3^{^{\prime}}}^{B}-i\left\vert \alpha\right\rangle _{3}^{A}\left\vert
\alpha\right\rangle _{2^{^{\prime}}}^{B}-e^{-i\phi_{33^{^{\prime}}}}\left\vert
\alpha\right\rangle _{3}^{A}\left\vert \alpha\right\rangle _{3^{^{\prime}}%
}^{B}\right)  ,\label{GFS}%
\end{align}
where $\phi_{22^{^{\prime}}}$, $\phi_{33^{^{\prime}}}$ are the relative phases
acquired by the lasers due to the gravitational interactions between the two
pulses along paths $2$ and $2^{^{\prime}}$, $3$ and $3^{^{\prime}}$,
respectively. The phases of the free evolution when the laser pulses move are
ignored since all paths have the same lengths. Other structures are also
feasible, e.g. to locate both L shapes one on top of the other or to locate
them as only branch $2$ is close to branch $2^{\prime}$, and branches $3$ and
$3^{\prime}$ are further away from each other, but the function is the same as
our suggested structure.

\begin{figure}[ptb]
\centering
\includegraphics[width=1\columnwidth]{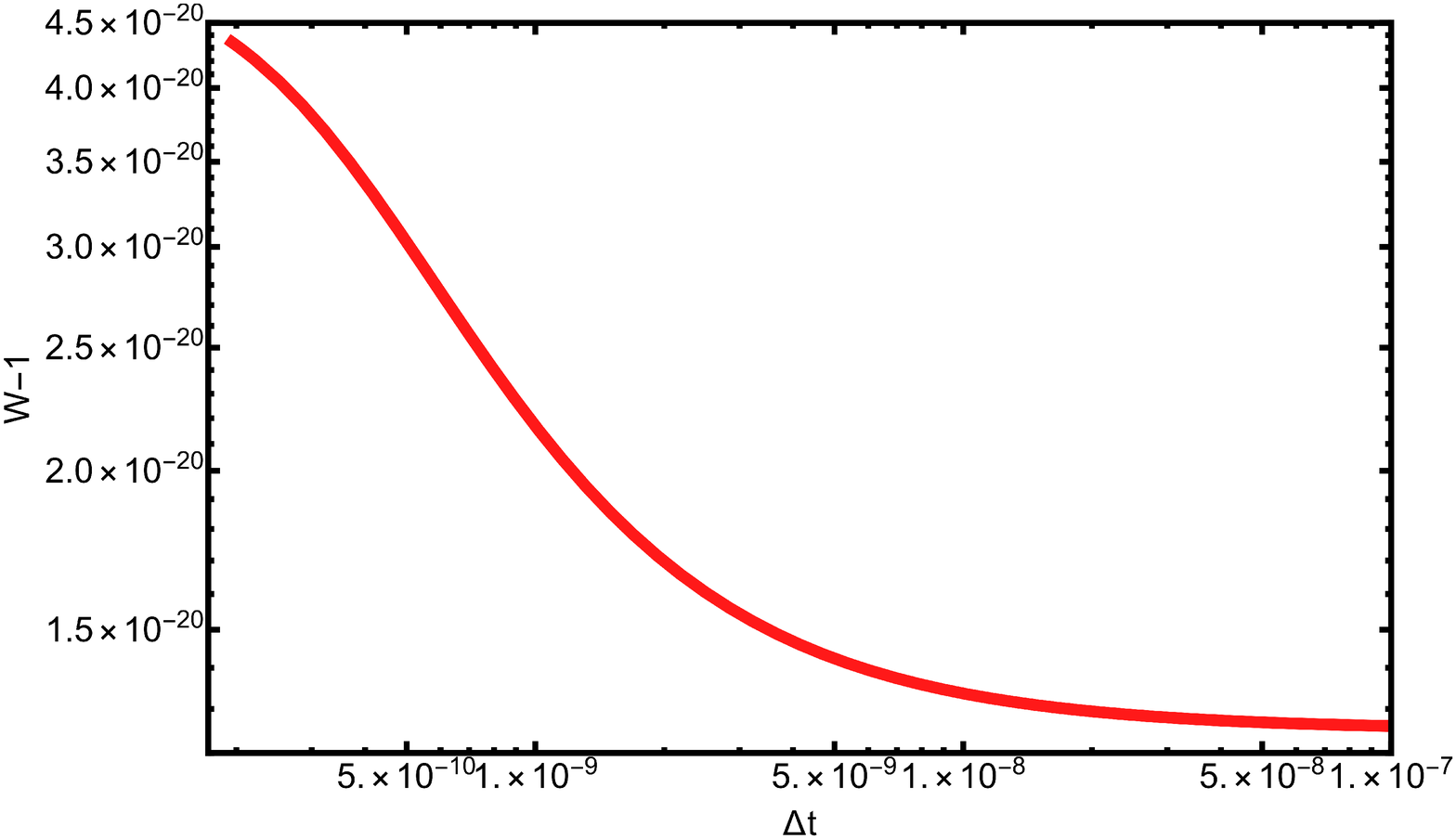} \caption{(Color online)
Log-log plot for entanglement witness W as a function of the emitted time
difference $\Delta t$ between two pulses $1$ and $1^{\prime}$ in Fig. 2. The
corresponding parameters are taken as $L=39$ m for the length of every path,
$d=0.9$ m for the separation of two interferometers, and $E_{laser}=339$ J for
the energy of the laser pulses. }%
\label{Fig3}%
\end{figure}

The relative phases $\phi_{22^{^{\prime}}}$ and $\phi_{33^{^{\prime}}}$ are
estimated according to the interaction Hamiltonian given by Eq. (\ref{HL2}),
and can be calculated approximately as
\begin{equation}
\phi_{22^{^{\prime}}}=\phi_{33^{^{\prime}}}\simeq\frac{{\hat{H}}_{int,r}^{G}%
t}{\hbar}=\frac{4Gm_{eff}^{2}L}{\hbar vd}+\frac{4Gm_{eff}^{2}L}{\sqrt
{d^{2}+\left(  v\Delta t\right)  ^{2}}\hbar v},\label{Int-Hamiltonian4}%
\end{equation}
where $t=2L/v$ and $2L$ is the distance of the paths that each pulse moves.
Taking the parameter for the length of the path as $L=39$ m, the velocity of
the pulse through the medium $v=0.9c$, the separation between two
interferometers $d=0.9$ m, the energy of the laser pulse $E_{laser}=339$ J
\cite{hmw11}, and the time difference for the emission of the two pulses
$\Delta t=0$, it is estimated that the relative phase $\approx10^{-11}$, which
is smaller than the current measurement sensitivity ($\sim10^{-8\text{ }}$for
the detection of Holometer \cite{hms}). Even so, the measured phase would
increase if the paths that the pulse moves are elongated, which can be done
using the Fabry--P\'{e}rot optical resonant cavity as in the measurement of
gravitational waves by LIGO \cite{rh00}. Compared with the other proposals
with massive particles \cite{bmm17,mv17}, it is advantageous in the
measurement sensitivity \cite{hms}. Moreover, in our suggestion, the retarded
effect can be shown, as in the following discussion about entanglement
generated through gravitational interaction between two pulses.

There are many methods to measure bipartite entanglement, such as von Newman
entropy, relative entropy, entanglement of formation, (logarithmic)
negativity, concurrence, entanglement witness and so on \cite{hhh09}. For our
discussed process, the concurrence or entanglement witness is proper. But the
calculation of concurrence requires full state tomography, so in the paper, we
adopt the entanglement witness \cite{hhh96,bmt00} to estimate the entanglement
in the quantum state (\ref{GFS}) for convenience. As discussed in Ref.
\cite{mmb20}, entanglement witness can measure entanglement between two pulses
and avoid the disturbance from the correlation between the pulse and its
environment. We use the entanglement witness as in Ref. \cite{bmm17,mmb20},
$W=\left\vert \left\langle \sigma_{x}^{(1)}\otimes\sigma_{z}^{(2)}%
\right\rangle -\left\langle \sigma_{y}^{(1)}\otimes\sigma_{y}^{(2)}%
\right\rangle \right\vert $ where $\sigma_{x}$, $\sigma_{y}$, and $\sigma_{z}$
are Pauli operators, and the superscripts (1) and (2) represent the two pulses
in different interferometers. With this definition, the witness has such a
property that it evaluates to greater than $1$ only if the state is entangled.
For our suggestion, two different paths are equivalent to two different spin
states. Thus we can calculate the witness using the state in Eq. (\ref{GFS})
and obtain%
\begin{equation}
W=\sqrt{5-4\cos\left(  \phi_{22^{^{\prime}}}+\phi_{33^{^{\prime}}}\right)
}=\sqrt{5-4\cos\left(  2\phi_{22^{^{\prime}}}\right)  },\label{ewr}%
\end{equation}
where in the last step, we take the relation $\phi_{22^{^{\prime}}}%
=\phi_{33^{^{\prime}}}$ according to the structure of Fig. 2. The result
(\ref{ewr}) is presented in Fig. 3, and it is seen that the entanglement is
decreasing with the increase of the retarded time, as expected. Moreover, we
give the relationship between the entanglement witness and the arm length in
Fig. 4, which shows that the generated entanglement will increase with the
increasing arm length.

\begin{figure}[ptb]
\centering
\includegraphics[width=1\columnwidth]{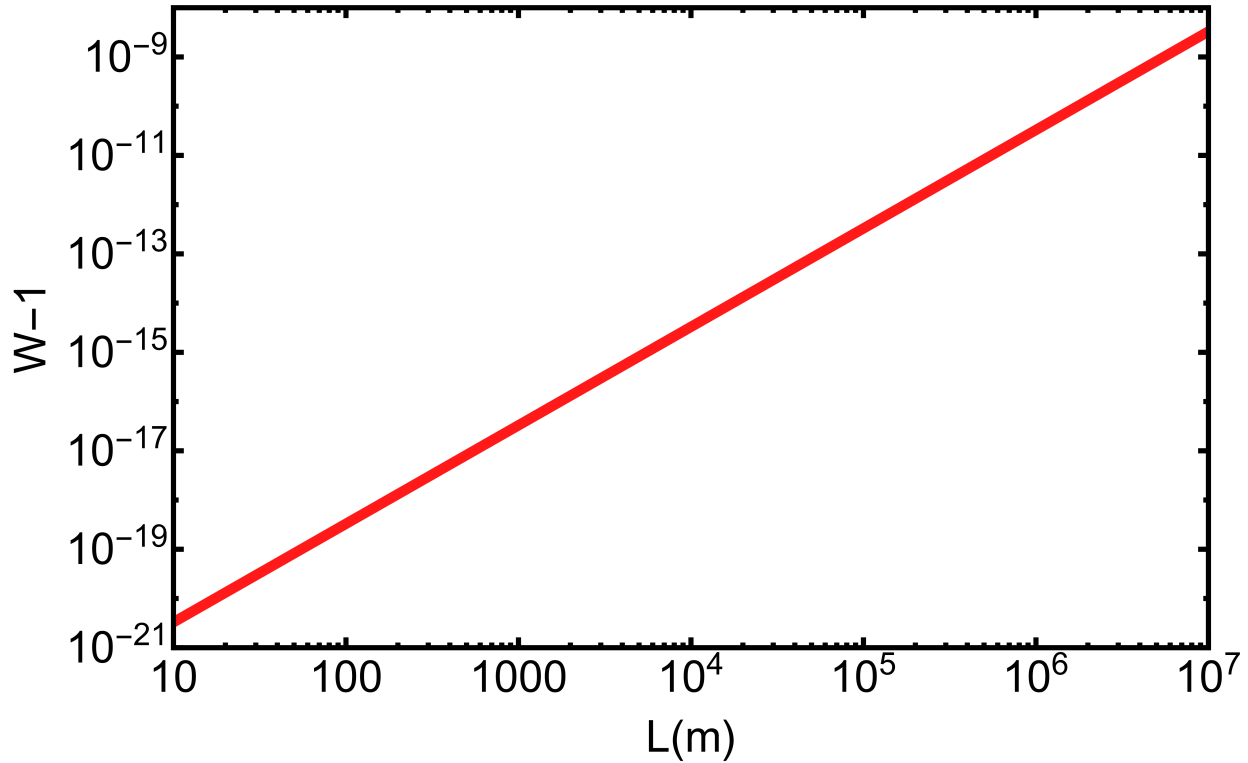} \caption{(Color online)
Log-log plot for entanglement witness W as a function of the armlength $L$.
The corresponding parameters are taken as the same values as in Fig. 3. }%
\label{Fig4}%
\end{figure}

We have presented the generation of entanglement above. Now an observable
scheme is given according to Fig. 2. Consider a setup consisting of two
interferometers $I_{A}$ and $I_{B}$, similar to the structure of
\textquotedblleft Holometer\textquotedblright\ \cite{hcj12}. Assume that the
two interferometers are placed horizontally at the same height and can be
linked only through gravitational interaction. Their interference results
would change if the two laser pulses could get entangled through gravitational
interaction in Eq. (15). The measurement for the generated entanglement is
related to the phase change, as seen in Eq. (\ref{ewr}). In the experimental
implementation, the electromagnetic leakage is a large disturbance for the
measurement, so the electromagnetic screen must be made so nice. Our earlier
estimation using the parameters of the Holometer \cite{hms} for the relative
phase ($\sim10^{-8\text{ }}$) includes the electromagnetic noise, so it can be
improved if the electromagnetic screen is better. With this phase sensitivity,
the minimum amount of entanglement witness for the measurement is $1+10^{-16}%
$. To detect the minimum entanglement witness, the arm length of the
interferometer has to reach the level of $10^{3}$ metres, which has been
realized in LIGO \cite{aaa16}, but whether it is proper to detect the
generated entanglement requires further analysis.

Gravitational waves have the quadrupole effect. That is, when the
gravitational perturbation generated by the interferometer $I_{A}$ influences
the interferometer $I_{B}$, one arm of $I_{B}$ should be lengthened and the
other one should be shortened by the proper placement of two interferometers.
But the gravitational-wave-like quadrupole effect is not presented clearly
here due to ignorance of interaction between the orthogonal paths (i.e.
between $2$ and $3^{^{\prime}}$, or $3$ and $2^{^{\prime}}$). The ignorance is
because the component of $h_{00}$ is larger than other components on one hand,
and the perturbation is dependent on the distance it propagates on the other
hand. Actually, the result of entanglement generation would not change even
after adding these considerations.

Finally, it has to be pointed out that the reason for the decrease of
entanglement in Fig. 3 with time difference is that the interaction between
two pulses would become weaker when the distance it propagates become larger,
as that for the gravitational waves.

\section{Conclusion}

In this paper, we have investigated the thought of using entanglement
generation through the gravitational interaction to show the quantum property
of the gravitation, but the gravitational interaction is generated by the
laser pulses that consist of massless particles instead of the earlier
suggested massive particles. At first, we review how the gravitational field
is generated when a pulse moves through a medium along the line of M. O.
Scully. It is noted that the generated gravitational field is time-varying. To
add the influence of propagating time for the metric perturbation, we have
considered the retarded solution for the interaction between two pulses. It
provides a self-consistent check for the required locality criterion that is
crucial for confirming the quantum gravitational field. Since the influence of
metric perturbation is dependent on the propagation time, only the interaction
between two parallel pulses is calculated. When the BSs are implemented for
two pulses, we have shown that entanglement can be generated, which provides
another possibility to show that the gravitational field is quantum. We have
also provided the detectable proposal using the Holometer-like structure of
two interferometers, in which the pulses in one interferometer can influence
the interference in another. Such a setup could induce entanglement by the
gravitational interaction between the two pulses. Although entanglement
generated by the gravitational perturbation is small, it can be detected in
the near future due to the rapid development of laser technology.

\section{Acknowledge}

We want to thank the reviewer for the critical and helpful comments. This work
is supported by NSFC (No. 11654001 and No. 91636213).

\section{Appendix1: The retarded effect}

Here we present the retarded effect of the gravitational interaction clearly.
Start with the linearized Einstein equation (\ref{leq2}). If time retardation
is not considered, the formal solution can be expressed as
\begin{equation}
h\left(  x,y,z,t\right)  =\frac{\kappa}{4\pi}\int_{-\infty}^{\infty}%
dx^{\prime}\int_{-\infty}^{\infty}dy^{\prime}\int_{-\infty}^{\infty}%
dz^{\prime}\frac{T_{00}\left(  x^{\prime},y^{\prime},z^{\prime},t\right)
}{\left[  \left(  x^{\prime}-x\right)  ^{2}+\left(  y^{\prime}-y\right)
^{2}+\left(  z^{\prime}-z\right)  ^{2}\right]  ^{1/2}},
\end{equation}
where $h\left(  x,y,z,t\right)  =h_{00}\left(  x,y,z,t\right)  $ is assumed
according to the text. When we consider the finite propagation of the
gravitational field, the retarded solution becomes,
\begin{align}
h_{r}\left(  x,y,z,t_{r}\right)   &  =\frac{\kappa}{4\pi}\int_{-\infty
}^{\infty}dx^{\prime}\int_{-\infty}^{\infty}dy^{\prime}\int_{-\infty}^{\infty
}dz^{\prime}\frac{T_{00}\left(  x^{\prime},y^{\prime},z^{\prime},t_{r}\right)
}{\left[  \left(  x^{\prime}-x\right)  ^{2}+\left(  y^{\prime}-y\right)
^{2}+\left(  z^{\prime}-z\right)  ^{2}\right]  ^{1/2}}\nonumber\\
&  =\frac{\kappa\varepsilon_{0}E_{0}^{2}A}{8\pi}\int_{-\infty}^{\infty
}dx^{\prime}\int_{-\infty}^{\infty}dy^{\prime}\int_{-\infty}^{\infty
}dz^{\prime}\frac{\left[  \theta\left(  v\left(  t_{r}+t_{d}\right)
-x^{\prime}\right)  -\theta\left(  vt_{r}-x^{\prime}\right)  \right]
\delta\left(  y^{\prime}\right)  \delta\left(  z^{\prime}\right)  }{\left[
\left(  x^{\prime}-x\right)  ^{2}+\left(  y^{\prime}-y\right)  ^{2}+\left(
z^{\prime}-z\right)  ^{2}\right]  ^{1/2}},\label{rsh}%
\end{align}
where $t_{r}=t-\sqrt{\left(  x^{\prime}-x\right)  ^{2}+\left(  y^{\prime
}-y\right)  ^{2}+\left(  z^{\prime}-z\right)  ^{2}}/v$ is the retarded time,
and the Eq. (\ref{see}) is used in the second line. To solve this further, we
make a coordinate transformation from $x^{\prime}$ to $\xi$ with the form%
\begin{equation}
\xi=\left(  x^{\prime}-x\right)  +\left[  \left(  x^{\prime}-x\right)
^{2}+\left(  y^{\prime}-y\right)  ^{2}+\left(  z^{\prime}-z\right)
^{2}\right]  ^{1/2}.
\end{equation}
This gives%
\begin{equation}
d\xi=\frac{\left(  x^{\prime}-x\right)  +\left[  \left(  x^{\prime}-x\right)
^{2}+\left(  y^{\prime}-y\right)  ^{2}+\left(  z^{\prime}-z\right)
^{2}\right]  ^{1/2}}{\left[  \left(  x^{\prime}-x\right)  ^{2}+\left(
y^{\prime}-y\right)  ^{2}+\left(  z^{\prime}-z\right)  ^{2}\right]  ^{1/2}%
}dx^{\prime}.\label{dtf}%
\end{equation}
Then, substituting the transformation (\ref{dtf}) into the Eq. (\ref{rsh}), we
have%
\begin{equation}
h_{r}\left(  x,y,z,t_{r}\right)  =\frac{\kappa\varepsilon_{0}E_{0}^{2}A}{8\pi
}\int_{\xi(a)}^{\xi(b)}d\xi\frac{\left[  \theta\left(  v\left(  t+t_{d}%
\right)  -x-\xi\right)  -\theta\left(  vt-x-\xi\right)  \right]  }{\xi},
\end{equation}
where the integral for the coordinates $y$ and $z$ has been made and the
retarded time is included in the variable $\xi$. $\xi(a)=\xi(vt)=\left(
vt-x\right)  +\left[  \left(  x^{\prime}-x\right)  ^{2}+\left(  y^{\prime
}-y\right)  ^{2}+\left(  z^{\prime}-z\right)  ^{2}\right]  ^{1/2}$ and
$\xi(b)=\xi(vt+vt_{d})=\left(  vt+vt_{d}-x\right)  +\left[  \left(
vt+vt_{d}-x\right)  ^{2}+\left(  y^{\prime}-y\right)  ^{2}+\left(  z^{\prime
}-z\right)  ^{2}\right]  ^{1/2}$, which is related to the length of the pulse.
Thus, the retarded solution is obtained as%
\begin{align}
h_{r}\left(  x,y,z,t_{r}\right)   &  =\frac{\kappa\varepsilon_{0}E_{0}^{2}%
A}{8\pi}\ln\left(  \frac{\xi(b)}{\xi(a)}\right)  \nonumber\\
&  =\frac{\kappa\varepsilon_{0}E_{0}^{2}A}{8\pi}\ln\left(  \frac{\left(
vt+vt_{d}-x\right)  +\left[  \left(  x-vt-vt_{d}\right)  ^{2}+\left(
y-y^{\prime}\right)  ^{2}+\left(  z-z^{\prime}\right)  ^{2}\right]  ^{1/2}%
}{\left(  vt-x\right)  +\left[  \left(  x-x^{\prime}\right)  ^{2}+\left(
y-y^{\prime}\right)  ^{2}+\left(  z-z^{\prime}\right)  ^{2}\right]  ^{1/2}%
}\right)  .
\end{align}
This is just Eq. (\ref{metric2}), but it presents the retarded effect clearly
in the calculation. The following calculations are the same as in the third
section. Thus, the result in Eq. (\ref{HL2}) implicitly includes the finite
propagation time for the gravitational interaction. So, other results after
that would include the finite propagation time implicitly.

\section{Appendix2: The derivation of the interaction quantum Hamiltonian}

In this appendix, we give the results in Eq. (\ref{Hamiltonian-laser}) from
Eq. (\ref{hamil-int}) along the line in Ref. \cite{mv18}. At first, the
quantized gravitational field is written in terms of the graviton creation and
annihilation operators $\hat{a}(k,\sigma)$, $\hat{a}^{\dagger}(k,\sigma)$, as
\begin{equation}
{\hat{h}}_{\mu\nu}\propto\sum_{\sigma}\int\frac{d^{3}k}{\sqrt{\omega_{k}}%
}\{\hat{a}(k,\sigma)\varepsilon_{\mu\nu}(k,\sigma)e^{ik_{\lambda x^{\lambda}}%
}+H.c.\},
\end{equation}
where $\varepsilon_{\mu\nu}$ is the polarization tensor, $\sigma$ indicates
two nonvanishing gravitational polarizations and $\omega_{k}$, $k$ represent
the frequency and wave number of the mode, respectively.

For simplicity, consider a single polarization and a discrete sum over the
relevant gravitational quantum modes. The Hamiltonian which involves two
masses and the gravitational field is
\begin{equation}
\hat{H}=m_{eff}c^{2}\left(  \hat{b}_{A}^{\dagger}\hat{b}_{A}+\hat{b}%
_{B}^{\dagger}\hat{b}_{B}\right)  +\sum_{k}\hbar\omega_{k}\hat{a}_{k}%
^{\dagger}\hat{a}_{k}-\sum_{k,n\rightarrow\{A,B\}}\hbar g_{k}\hat{b}%
_{n}^{\dagger}\hat{b}_{n}\left(  \hat{a}_{k}e^{ikx_{n}}+\hat{a}_{k}^{\dagger
}e^{-ikx_{n}}\right)  ,
\end{equation}
where the first two terms are the free Hamiltonian of the lasers and the
field, respectively. The third term is the interaction Hamiltonian which
derives from the expression $\hat{H}_{int}=-\frac{1}{2}{\hat{h}}_{\mu\nu}%
{\hat{T}}^{\mu\nu}$. Assumed that the gravitation-matter coupling constant is
given by $g_{k}=2mc\sqrt{\frac{2\pi G}{\hbar\omega_{k}V}}$, in which the
prefactor $2$ is due to the two similar terms in Eq. (\ref{hamil-int}). The
evolution of two lasers of value $m_{eff}$ at position $x_{A}$ and $x_{B}$
interacting with the initial gravitational vacuum state can be solved
exactly,
\begin{equation}
e^{iHt}\left\vert \alpha\right\rangle \left\vert \alpha\right\rangle
\left\vert 0\right\rangle =\exp{[\hbar\sum_{k}V(k)t]}\left\vert \alpha
\right\rangle \left\vert \alpha\right\rangle {\sum_{k}}\frac{g_{k}}{\omega
_{k}}\left(  e^{-ikx_{A}}+e^{ikx_{B}}\right)  ,
\end{equation}
where $\left\vert \alpha\right\rangle $ is the coherent state describing the
laser, $\left\vert 0\right\rangle $ represents the gravitational vacuum state,
and $V(k)=\frac{g_{k}^{2}}{2\omega_{k}}\left(  1+2\cos\left(  -ik(|x_{B}%
-x_{A}|)\right)  \right)  $.

Using the integral to replace the sum over $k$ and noting that only the
position-dependent part of $V(k)$ contributes to the phase difference, we
obtain
\begin{equation}
Re\{V\int dk\frac{16\pi Gm_{eff}^{2}}{\hbar k^{2}V}e^{-ik|x_{B}-x_{A}%
|}\}=\frac{4Gm_{eff}^{2}}{\hbar D}, \label{qa}%
\end{equation}
which gives the interaction phase with $\left\vert x_{B}-x_{A}\right\vert =D$
for $x_{B}\rightarrow0$. According to our suggested effective mass
$m_{eff}=\rho V/c^{2}$, Eq. (\ref{Hamiltonian-laser}) is restored from Eq.
(\ref{qa}).

\section{Data Availability Statement}

The authors declare that the data supporting the findings of this study are
available within the article.


\begin{thebibliography}{99}
\bibitem {hms}The sensitivity of the phase for the Holometer (see A. S. Chou,
R. Gustafson, C. Hogan C, et al., Phys. Rev. D \textbf{95}, 063002 (2017)) is
estimated using the relation, $\left(  \delta\phi\right)  ^{2}\sim\frac
{hc^{2}}{2PL\lambda}$, where $h$ is the Planck constant, $P$ is the optical
power of the light with $P=2000$ W, $\lambda$\ is the wavelength of the light
with $\lambda=1064$ nm, and $L$ is the arm length of the interferometer with
$L=39$ m. In the laser interferometry, the phase sensitivity can reach at
$10^{-8}$, while in the experiments that used massive particles, the phase
sensitivity is $0.01$, for mass $m=10^{-15}$ kg.
\end{thebibliography}
\end{document}